# Employing per-component time step in DSMC simulations of disparate mass and cross-section gas mixtures

Roman V. Maltsev[1]

*Address: 8-22, Koltsovo, Novosibirsk, 630559, Russia. Independent scientist.*

**Abstract.** A new approach to simulation of stationary flows by Direct Simulation Monte Carlo method is proposed. The idea is to specify an individual time step for each component of a gas mixture. The approach consists of modifications mainly to collision phase and recommendation on choosing time step ratios. It allows softening the demands on the computational resources for cases of disparate collision diameters of molecules and/or disparate molecular masses. These are the cases important in vacuum deposition technologies. Few tests of the new approach are made. Finally, the usage of new approach is demonstrated on a problem of silver nanocluster diffusion in carrier gas argon in conditions of silver deposition experiments.

## 1 Introduction

**The Direct simulation Monte Carlo (DSMC) method** [1] is a standard for simulation of nonequilibrium rarefied gas flows described by Boltzmann equation. This method is based on tracking individual molecules (simulators), considering them moving independently with occasional discrete events of pair collisions applied according to a statistical model. After the steady state is reached, the average of simulator parameters over many time steps derives macroparameters of the flow. The method has three discretization parameters.

---
[1] Contact e-mail: rmaltsev@gmail.com

First, the displacement and collision phases are uncoupled to speed up a computation, this gives the time step parameter $\Delta t$.

Second, a considerable finite number of simulators are used to simulate a large number of physical molecules. The symmetry of Boltzmann equation allows reducing the number density $n$ of molecules by increasing the collision cross-section $\sigma_T$, keeping the local mean free path $\lambda$ unchanged:

$$n \to \frac{n}{F}, \quad \sigma_T \to F\sigma_T, \quad \lambda = \frac{1}{\sqrt{2}n\sigma_T},$$

here $F$ is the number of physical molecules represented by a simulator.

Third, for a finite number of simulators, collisions can no longer be a point, as it should in Boltzmann equation, so, the simulation area is divided into cells of linear size $h$, and collision partners are chosen randomly within a same cell, causing spatial collision separation.

As is said, DSMC is the most practical method for numerical simulation of nonequilibrium rarefied gas flows. However, statistical nature of DSMC forces to calculate a lot of time steps to collect enough samples to get good statistical averages, as the noise amplitude is inversely proportional to the square root of the simulated period of time. When the flow is close to equilibrium, gradients of macroparameters are too small to resolute. This forces researches to use modified DSMC [2-5] or even alternative [6-7] methods to solve Boltzmann equations.

**Discretization** causes distortion of flow parameters and effective transport coefficients, error is a second order on cell size and time step, i.e. $\sim \left(\frac{h}{\lambda}\right)^2$ [8], and $\sim(\nu_C \Delta t)^2$ [9] ($\nu_C$ is local mean collision frequency), and first order on $F$, i.e. $\sim \frac{h}{\lambda \overline{N}}$ [10] ($\overline{N}$ is the mean number of simulators in a cell, and $\lambda \overline{N}$ is invariant on local density). Simulating $d$-dimensional flow of characteristic linear size $L$ and characteristic mean free path $\lambda$ keeping the same distortion requires $\sim \left(\frac{L}{\lambda}\right)^d$ cells and proportional number of simulators, these determine demands on both the computer memory and number of operations per time step. With the increase of flow density, time step has to shrink, while relaxation of the flow slows down, therefore, the number of time steps to reach steady state is $\sim \left(\frac{L}{\lambda}\right)^2$. Sometimes reaching a steady state is more costly than collecting a good average of flow properties. When detection of steady state is not obvious, researchers have to use convergence detection algorithms [11].

**Majorant collision frequency** (MCF) scheme [12] assumes each possible pair $(i,j)$ of simulators in a cell has its own collision frequency:

$$v_{ij} = \frac{F \sigma_T c_r}{V_C}, \qquad (1.1)$$

here $c_r$ is relative velocity and $V_C$ is the cell size. The algorithm to accomplish this is as following. First, a value of majorant collision frequency $v_{max}$ is chosen in each cell:

$$v_{max} = \frac{F \{\sigma_T c_r\}_{max}}{V_C}.$$

The value $v_{max}$ is to be high enough, so, only negligible fraction of pairs exceeds it. In practice, local value of $v_{max}$ is updated each time it is exceeded. Majorant frequency is used to calculate the expected number of virtual collisions:

$$\overline{K_{max}} = \frac{N(N-1)}{2} v_{max} \Delta t,$$

here $N$ is an instantaneous number of simulators in a cell. The actual number $K_{max}$ of virtual collisions is a sample from Poisson distribution with mean value $\overline{K_{max}}$. Then, $K_{max}$ random collision pairs are chosen within a cell, each virtual collision is accepted (and simulated) with probability $v_{ij}/v_{max}$.

In contrast to original No Time Counter (NTC) scheme [1], a priori knowledge of mean number $\overline{N}$ of simulator in a cell is not required and do not need to be evaluated on-the-fly. This makes MCF scheme insensitive to $\bar{N}$ [10], while original NTC scheme slightly is [13]. In latter version of NTC scheme, Bird proposed to use the same $\overline{K_{max}}$ as in MCF scheme, but with $K_{max}$ calculated as $\overline{K_{max}}$ randomized to one of two nearest integers, rather than sampling from Poisson distribution.

The usage of MCF scheme is supposed everywhere in this paper, though everything might still work for NTC scheme.

**Component weights.** Sometimes, the concentration of some component is too small to collect good averages. A solution is to use smaller value of $F$ (simulator weight) for scanty component only to increase a number of simulators reasonably. Using small $F$ for all components together would increase number of simulators enormously. However, collision scheme must be modified [14] for colliding simulators of different weights. Eq. (1.1) then becomes:

$$v_{ij} = \frac{\max\{F_i,F_j\}\sigma_T c_r}{V_C}, \qquad (1.2)$$

here $F_i$ is component weight of simulator $i$. If the collision is accepted, simulator $i$ changes its parameters to post-collision ones with the probability $\frac{F_i}{\max\{F_i,F_j\}}$. Hence, a simulator with smaller weight always receives post-collision parameters, another one may either receive post-collision or retain pre-collision parameters. This non-symmetry brings a number of drawbacks. First, the scheme is non-conservative, so, random walks of momentum and energy are present. Second, the simulator with smaller weight may collide with same pre-collision simulator of greater weight few times in a row. This causes specific distortion of distribution function[2]. This distortion decreases as more simulators (especially for component of greater weight) are used, or as component weights become less distinct. If the component of smaller weight also has much greater mass, it is severely sensitive to this distortion.

**Spatial weights.** A simulator weight $F$ can depend on coordinates. However, simulators have to be duplicated/decimated when they travel between regions with different $F$. Spatial weights retain all drawbacks of component weights. Additionally, randomization in duplication/decimation scheme causes random walks of density, sometimes severe [15]. Nevertheless, particular case of radial weights is almost indispensable for axisymmetric flows. Otherwise, there are too few simulators near the axis and too many on peripheries.

---

[2] Explicit rejection of such collisions helps, but does not solve the problem of specific distortion completely.

**Kannenberg's approach.** In some flows, the density can vary by orders of magnitude. Lower density regions may be correctly simulated with larger cells, longer time step, and larger $F$, i.e. fewer simulators[3]. Kannenberg proposed an approach [16] to take advantage of this, though it works only for stationary flows. Let introduce the time step scaling factor $T$ in each simulation cell, with local time step becoming $T\Delta t$, while now $\Delta t$ is the reference global time step. Local time step is used in both transposition and collision phases. In transposition phase, when a simulator moves to a cell with different corresponding $T$, it retains the unused fraction of time step, i.e. absolute residue of time step adjusts to new value of $T$. Additionally, local values of $F$ are scaled in proportion to $T$, so, the ratio $W = \frac{F}{T}$ retains the same in all cells ($W$ will be called the combined weight later on). Duplication/decimation is not needed in the case $W = const$, as simulators tend to crowd more in cells of smaller time step in natural way. The collision frequency of a pair (1.1) with respect to the global time step now may be written in the form:

$$v_{ij} = T^2 \frac{W \sigma_T c_r}{V_C}, \qquad (1.3)$$

where the first incarnation of $T$ converts $W$ to $F$, and second one scales the global time step.

To apply the approach, first, larger cells are to be chosen in the region of lower density. Then, local $F$ and $T$ are increased. Amount of increase is limited by allowing neither $\frac{h}{\lambda \overline{N}}$, nor $v_C T \Delta t$ discretization indexes exceed their preferred values.

Adapting local time step gives triple advantage. First, it decreases a total number of simulators required. Second, it locally accelerates the convergence toward the steady state in low-density regions. Third, decreasing number of simulators does not harm the statistical convergence, as intra-cell correlations now decay faster, balancing the reduction in number of simulators. All these decrease demands on computational resources without detriment in precision.

---
[3] Not in 1D problems.

Despite symmetrical collisions, the approach is not strictly conservative, as the mass, momentum and energy may redistribute among simulators of different $F$. This causes inconvenience when simulating closed flows. Nevertheless, after the flow achieves steady state, those invariants stabilize and oscillate around established values, and do not suffer from random walks typical for non-conservative spatial weights. Thus, the approach may be called quasi-conservative.

**Repeated collisions.** The distortion caused by limited number of simulators is interrelated [17] with the probability of repeated collisions (i.e. two simulators collide twice in a row without colliding other simulators in between), thus, the fraction of repeated collisions may serve as measure of statistical dependence between simulators, which spoils the assumption of molecular chaos. In [10], an estimation of this probability $p_{rpt}$ is proposed, that yields $\sim \frac{h}{\lambda \overline{N}}$ discretization index in simplest case:

$$p_{rpt} \sim \overline{v_{ij} t_P} \sim \frac{F \overline{\sigma_T c_r}}{V_C} \cdot t_P,$$

$$t_P \sim \frac{h}{\overline{c_r}},$$

$$p_{rpt} \sim \frac{F \overline{\sigma_T c_r}}{V_C} \cdot \frac{h}{\overline{c_r}} \sim \frac{F \overline{\sigma_T} h}{V_C}, \tag{1.4}$$

$$\frac{F \overline{\sigma_T} h}{V_C} = \frac{n \overline{\sigma_T} h}{n V_C/F} = \frac{\sqrt{2} n \overline{\sigma_T} h}{\sqrt{2} \cdot \overline{N}} = \frac{1}{\sqrt{2}} \cdot \frac{h}{\lambda \overline{N}},$$

Here $t_P$ is an estimate of the time a pair of just collided simulators needs to separate far enough not to gather in a same cell and collide again. It is assumed $\Delta t < t_P$, i.e. simulators don't separate faster than 1-2 cells per time step. Simulators need at least one time step to separate, so, with $\Delta t > t_P$, an effective value of $t_P$ becomes $\Delta t$. In result, $p_{rpt}$ rises linearly with $\Delta t$ [10], and the distortion caused by the limited number of simulators increases. Thus, very small cells should be used together with small enough time step. This is important for 1D problems, when $\frac{h}{\lambda \overline{N}}$ is invariant on cell size.

For gas mixtures, $p_{rpt}$ is different for collisions between different components, being higher for collisions with bigger cross-section. In the case when component weights are used, $p_{rpt}$ is different for two simulators from a pair. For component $A$, the weight $F_B$ of component $B$ should be used to estimate $p_{rpt}$ in $A-B$ collisions, and vice-versa. Thus, component weights are useful not only to amplify scarce component, but also to use more simulators only for components with bigger collision diameter.

**Per-component time step** approach introduced in this paper is a new tool useful to simulate flows where each component has different characteristic time. For example, consider a binary gas mixture of light carrier component $A$ and small amount of large polyatomic component $B$ (a typical case in supersonic deposition experiments [18], when deposited precursor molecules need to be accelerated to high velocity), $A-B$ collision cross-section is usually much larger than $A-A$ one. As a result, $B$ has much greater collision rate than $A$, sometimes forcing to decrease single global time step (especially if $B$ consists mostly of light atoms). Additionally, as was shown above, large $A-B$ collision cross-section forces to use more simulators for $A$, otherwise $B$ will suffer from frequent repeated collisions with $A$. This is true even if concentration of admixture $B$ is small enough not to disturb the flow of carrier gas $A$, despite the fact that simulating the carrier $A$ alone may be done with less number of simulators.

Fortunately, the new approach allows to decrease only the time step for component $B$. Besides, less simulators of component $A$ are required when component $B$ is slowed down. This new approach works only for stationary flows.

## 2   Per-component time step

Let us start by sketching the Boltzmann equations for binary mixture of components A and B:

$$\begin{cases} \dfrac{1}{T_A} \cdot \dfrac{\partial f_A}{\partial t} + v_i \dfrac{\partial f_A}{\partial x_i} = F_A \cdot I_{AA}[f_A, f_A] + F_B \cdot I_{AB}[f_A, f_B] \\ \dfrac{1}{T_B} \cdot \dfrac{\partial f_B}{\partial t} + v_i \dfrac{\partial f_B}{\partial x_i} = F_B \cdot I_{BB}[f_B, f_B] + F_A \cdot I_{BA}[f_B, f_A] \end{cases}$$

Here $f_A$ and $f_B$ are distribution function of simulators for components $A$ and $B$, dependent on velocity vector $v_i$ and coordinates $x_i$. Functionals $I_{AA}, I_{AB}, I_{BA}, I_{BB}$ are standard collision integrals. $F_A$ and $F_B$ are component weights, i.e. number of real molecules per simulator. $T_A$ and $T_B$ are time scaling factors, which should both be equal to unity for normal Boltzmann equation. However, since a steady state is achieved, temporal derivations become zero, and equations are satisfied with arbitrary values of time scaling factors. Equations may be written in different form:

$$\begin{cases} \dfrac{\partial f_A}{\partial t} + (T_A v_i) \dfrac{\partial f_A}{\partial x_i} = T_A^{\;2} \cdot W_A \cdot I_{AA}[f_A, f_A] + T_A T_B \cdot W_B \cdot I_{AB}[f_A, f_B] \\ \dfrac{\partial f_B}{\partial t} + (T_B v_i) \dfrac{\partial f_B}{\partial x_i} = T_B^{\;2} \cdot W_B \cdot I_{BB}[f_B, f_B] + T_A T_B \cdot W_A \cdot I_{BA}[f_B, f_A] \end{cases}$$

$$W_A = \dfrac{F_A}{T_A}, \qquad W_B = \dfrac{F_B}{T_B}.$$

This form gives a hint about modifying the simulation scheme. First, the transposition phase should scale the global time step with given time step scaling factor $T$, just like in Kannenberg's approach. Second, collisions between simulators of the same component may be treated the same way as in Kannenberg's approach as well (1.3). Third, more generalized version of (1.2) for inter-component collisions should be used:

$$v_{ij} = T_A T_B \dfrac{\max\{W_A, W_B\} \sigma_T c_r}{V_C}.$$

Again, if two components have different combined weights (i.e. $W_A \neq W_B$), the one with smaller combined weight always receives post-collision parameters, and another does with probability $\dfrac{\min\{W_A, W_B\}}{\max\{W_A, W_B\}}$. If combined weights are equal, both simulators always change their parameters in a collision; hence, the collision scheme is symmetrical and quasi-conservative in this case.

As in Kannenberg's scheme, time step scaling factors may change from cell to cell – moreover, now time step scaling factors of various components may change independently, adapting to local characteristic times.

When spatial weights are used as well, so that either $F$ and/or $W$ includes explicit dependence on coordinates, the simulator duplication/decimation should be applied when combined weight $W$ at the end of displacement trajectory is different to one at the beginning. Note, the ratio of old and new $W$, not the ratio of $F$, should be used to calculate the expected coefficient of multiplication duplication/decimation procedure.

When collecting samples inside cells by "taking photo" (summing parameters of simulators inside), the component weight $F$ should be used to scale the contribution of a simulator. When collecting samples on simulators hitting or crossing some surface or boundary, as well as determining the number of simulators to inject from boundaries, the combined weight $W$ should be used instead.

Now let examine repeated collision probability and extend (1.4) to the case when different time steps for components $A$ and $B$ are used. Again, let $A$ be the light carrier gas with the mass $m_A$, and $B$ be the heavy admixture with the mass $m_B = \mathcal{M} \cdot m_A$, with $\mathcal{M} \gg 1$ being the mass ratio. The probability of repeated collision for component $B$ is:

$$p_{rpt\,B} \sim \frac{F_A \sigma_{AB} h}{V_C} \cdot \left\langle \frac{T_B \cdot \|\vec{v_A} - \vec{v_B}\|}{\|\vec{v_A} T_A - \vec{v_B} T_B\|_D} \right\rangle. \tag{2.1}$$

For component $A$, just swap indexes $A$ and $B$. The index $D$ means that only those vector components that lie in computational domain subspace should be taken into account (i.e. only one in 1D case, 2 in 2D case, and all 3 in 3D case). The worst case is when $\|\vec{v_A} T_A - \vec{v_B} T_B\|_D \to 0$, though it is unlikely and may be avoided by choosing slightly different time step scaling factors when flow predisposition to adverse component velocity ratio is expected.

Let examine the simplest case, when mean velocities may be supposed to be zero, and only thermal velocities contribute. Thermal velocities in equilibrium are in proportion to reciprocal square root of molecular mass. Eq. (2.1) and its version for $A$ then become:

$$p_{rpt\,B} \sim \frac{F_A \sigma_{AB} h}{V_C} \cdot \sqrt{\frac{1+\frac{1}{\mathcal{M}}}{\frac{1}{\mathcal{M}}+\left(\frac{T_A}{T_B}\right)^2}} = \frac{T_B W_A \sigma_{AB} h}{V_C} \sqrt{\frac{1+\frac{1}{\mathcal{M}}}{1+\frac{1}{\mathcal{M}}\cdot\left(\frac{T_B}{T_A}\right)^2}}, \tag{2.2}$$

$$p_{rpt\,A} \sim \frac{F_B \sigma_{AB} h}{V_C} \cdot \sqrt{\frac{1+\frac{1}{\mathcal{M}}}{1+\frac{1}{\mathcal{M}}\cdot\left(\frac{T_B}{T_A}\right)^2}} = \frac{T_A W_B \sigma_{AB} h}{V_C} \sqrt{\frac{1+\frac{1}{\mathcal{M}}}{\frac{1}{\mathcal{M}}+\left(\frac{T_A}{T_B}\right)^2}}. \tag{2.3}$$

As seen from (2.2), slowing down heavy component $B$ in respect to light $A$, favors the decrease of repeated collisions for $B$. Contrary, hastening $B$ causes more repeated collisions, up to $\sqrt{\mathcal{M}+1}$ times. Slowing down $A$ in respect to $B$, but keeping the same $W_A$ (i.e. balancing the slow-down by proportional increasing the number of simulators), is safe and does not cause more repeated collisions.

Now the light component (2.3). Both slowing and hastening $A$ is safe. Slowing down $B$ is safe as well. Hastening component $B$ while keeping the same $W_B$ (i.e. proportionally decreasing its number of simulators) causes more repeated collisions for $A$, up to $\sqrt{\mathcal{M}+1}$ times.

Thus, after combining both analyses, slowing down heavy component is safe in respect to $A-B$ collisions, while slowing down light component is safe only when balanced by the proportional increase in its number of simulators. Hastening light component is safe and allows decreasing its number of simulators proportionally. Hastening heavy component is unsafe and not recommended.

Effect of time scaling factors and combined weights on $A-A$ and $B-B$ collisions should not be overlooked as well and must be checked separately. Also, simulators should not separate by too many cells per time step. With different time steps for two components, it is now more probable in supersonic flows.

Though (2.2) and (2.3) are derived for zero mean velocities and equilibrium only, I use them as universal hint in my simulations and it seems to work well. Further computational investigations of the effect of selection of time step scaling factors are encouraged.

## 3  A test of slowing down the light component

A good model problem for the test is classic 1D heat transfer problem: a gas is contained between two parallel plates of different temperature, molecules experience full accommodation of momentum and energy on both plates, heat flux between plates is measured. It is well-known that even small admixture of a light gas to a heavy one may noticeably increase the heat conductivity of a mixture because of faster thermal speed of light component. For the test, let's take a model mixture of 10% He + 90% Xe. The heat conductivity of such a mixture is ≈1.5 times higher than of pure xenon. Helium thermal speed is ≈6 times higher, and collision rate is ≈2 times higher as well, in comparison with xenon.

The test will consist of simulating the heat transfer problem with three different time step factors $T_H$ for helium: 1 (standard DSMC), 1/1.7, and 1/3. Slowing down helium is balanced by the proportional increase in the number of simulators for it (keeping combined weight constant), thereby, quasi-conservative mode is provided. Wall heat flux $q$ is determined by sampling all the simulators colliding walls, and dependence of heat flux deviation on the global time step is to be studied. Variable Soft Spheres (VSS) collision model is used with Bird's parameters [1] for all three types of collisions. Temperatures of two walls are set 137 K and 536 K – fourfold difference (former one is lower than freezing point 161 K of real xenon, but it's not a disaster for a numeric test). The distance between walls and gas density correspond to the chosen value of similarity parameter:

$\mathrm{Kn}_0 = \frac{1}{\sqrt{2}n_0 \sigma_{H-H} L} = 0.1$, here $n_0$ is a total number density of the mixture, $\sigma_{H-H}$ is helium-helium collision cross-section, both determined at isothermic initial condition (273 K), $L$ is the distance between walls. The distance is divided by 500 uniform cells. The number of simulators are: 5000 for $T_H = 1$, 5350 for $T_H = 1/1.7$, and 6000 for $T_H = 1/3$.

Results of three series of simulation are shown in Fig. 1. The unit of time equals: $\frac{L}{753\ m/s}$. The "reference" value of heat flux $q_{ref}$ was determined using time step of 0.001, the deviation is calculated as: $\left|\frac{q}{q_{ref}} - 1\right|$. Actual points acquired by simulations are shown together with fitted curves. One can see, slowing down helium 1.7 times allows increasing global time step by 25% keeping the deviation under 1%, at the cost of 7% more total simulators, while keeping the same global time step decreases the heat flux deviation by 25%. Slowing down helium 3 times allows increasing global time step by 50%, at the cost of 20% more simulators, or, decreases deviation by 35% keeping the same global time step.

This test confirms that slowing down light admixture component may improve the precision of simulation. It keeps true down to minor time steps, where the deviation cannot be correctly resolved.

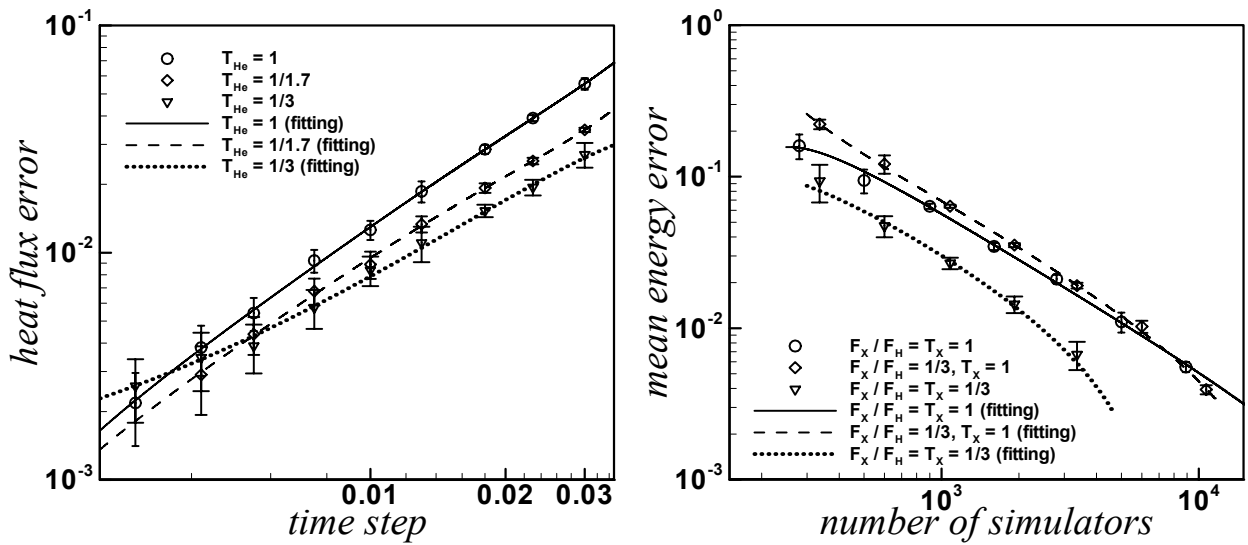

Fig. 1. Dependence of heat flux error on global time step, with different time step scaling of light component (helium). Actual points and fitted curves.

Fig. 2. Dependence of mean heavy component (xenon) energy after passing "compressed layer" on the total number of simulators, for different time step scaling of heavy component. Actual points and fitted curves.

## 4 A test of slowing down the heavy component

For this test, a good model problem is a supersonic flow of a mixture around an obstacle, with shock wave and compressed layer formed in front of an obstacle, and heavy component decelerating in compressed layer (because of disparate masses, relaxation between light and heavy molecules needs many collisions). Such a problem needs at least 2D simulation domain. However, 1D problem are simpler, have less parameters and allow to collect better statistics. The simplified 1D problem is following. Consider a gas mixture confined between two parallel plates, which model bounds of a compressed layer, with left one being a "shock wave", and right one a "target". Light component is diffusively reflected from both walls with full accommodation of energy and momentum. Heavy component reflects diffusively from left plane only. If a heavy molecule hits the right wall, it is considered as "deposited" and leaves the computation domain. As soon as it happens, a new heavy molecule is reemitted from the left bound with high velocity, being considered as just "entered" the compressed layer from supersonic jet. Thus, the number of molecules in simulation domain is constant. The mean energy $E$ of heavy molecules reaching right plate is studied.

Both plates are kept at 273 K. The mixture is 90% He + 10% Xe. Again, VSS model with parameters [1] is used. The similarity parameter: $\text{Kn}_0 = \frac{1}{\sqrt{2} n_0 \sigma_{H-H} L} = 0.04$. Xenon molecules are injected with the zero temperature and the velocity of 1684 m/s. The distance between plates is divided by 500 uniform cells. The global time step equals $0.001 \cdot \frac{L}{753 \, m/s}$. The varied parameter is total number of simulators, by the means of changing the combined weight. Three series of simulations are made: xenon time step scaling factor $T_X = 1$ and ratio of component weights $\frac{F_X}{F_H} = 1$ (standard DSMC), $\frac{F_X}{F_H} = T_X = 1/3$ (xenon is slowed down, its number of simulators is increased proportionally), and $\frac{F_X}{F_H} = 1/3$ with $T_X = 1$ (number of xenon simulators is increased, but time step is the same, non-conservative mode). In latter two cases, 20% more total simulators are used. The "reference" energy $E_{ref}$ is computed with standard DSMC using 300 000 simulators. Then, the deviation is computed as follows: $\left|\frac{E}{E_{ref}} - 1\right|$

Results are plotted in Fig. 2. One can see that when different time steps are used, the mean energy error decreases more than twice compared to standard DSMC. Or, the same error may be achieved using twice less simulators. Using just different component weights, without time step scaling, does not give such an improvement of precision. This is predictable and testifies that improvement is caused by using different time steps, not by changing the portion of xenon simulators.

The test confirms that slowing down heavy admixture component may improve the precision of simulation as well.

## 5  A qualitative test in 2D

One more test of slowing down the heavy component, but full 2D simulations are used this time. Again, let us choose the problem of the deceleration of heavy admixture in the compressed layer formed by the carrier gas in front of the flat plate [19], well-known to the author. However, this time let's just do the qualitative comparison and do not analyze the behavior of some error value. For this to be possible, let's advisedly use bare number of simulators.

The problem setting for the test is as follows. Consider the supersonic plane-parallel flow of a gas mixture, consisting of 95% He + 5% Xe, with temperature of 29.25 K and the speed of $V_\infty$ =1592 m/s. This corresponds to the stagnation temperature of 660 K and the Mach number of 8.05. The flat plate of the width $L$ is placed transversely to the flow. Plate temperature is 273 K, both components diffusively scatter from both sides of the plate, with full accommodation. The characteristic Knudsen number: $\text{Kn}_\infty = \frac{1}{\sqrt{2} n_\infty \sigma_{H-H} L} = 1$. Here $n_\infty$ is the total number density in undisturbed flow, $\sigma_{H-H}$ is helium-helium collision cross-section at 273 K (collision diameter is 2.3 Å).

The energy spectrum of heavy molecules colliding the front side of the plate is observed. Typical energy spectrum in these conditions consists of three contributions: the high-energy part of xenon molecules came from undisturbed flow and decelerated by multiple collisions with helium in compressed layer, the low-energy part of xenon molecules collided with the plate before and then with helium, and middle-energy part of xenon molecules came from collisions between low-energy and high-energy xenon.

The cell size is $h = 0.025 \cdot L$. The number of simulators is defined by: $\frac{\sqrt{2} F_H \sigma_{H-H}}{h} = 0.5$. Four simulations are made: standard DSMC ($\frac{F_X}{F_H} = T_X = 1$), DSMC with component weights ($\frac{F_X}{F_H} = 0.15$), new approach in quasi-conservative mode ($\frac{F_X}{F_H} = T_X = 0.15$), and the reference simulation – standard DSMC with 25x number of simulators.

Fig. 3 presents the high-energy part of energy spectrums for these cases. The unit of energy is $\frac{m_X V_\infty^2}{2}$, with $m_X$ being a mass of xenon (131 a.u.), the integral flux density is normalized to the flux density $n_X V_\infty$ in undisturbed flow, with $n_X$ being a number density of xenon. One can see, standard DSMC shows the most deviation from the reference spectrum, the new approach – the least deviation, the DSMC with component weights is better, than standard DSMC, but worse, than new approach. In latter case, using the component weights for xenon improves the simulation of xenon-xenon collisions, but xenon-helium collisions are of same quality as in standard DSMC. In contrast, new approach improves xenon-helium collisions as well as xenon-xenon, though it needs more time steps to collect good averages.

Fig. 4 shows the temperature of xenon. This temperature has high values (over 3500 K) and is non-physical because of high non-equilibrium in the rarefied compressed layer, but it still characterizes the thermal energy of xenon, besides, some high-energy xenon-xenon collisions do happen. Again, the new approach gives the closest values. At $X/L < -3$, the non-conservative approach of component weights shows its usual artifact of overestimating the temperature of amplified heavy component.

The test confirms that the new approach can improve complex 2D simulations of disparate mass gas mixture flows as well.

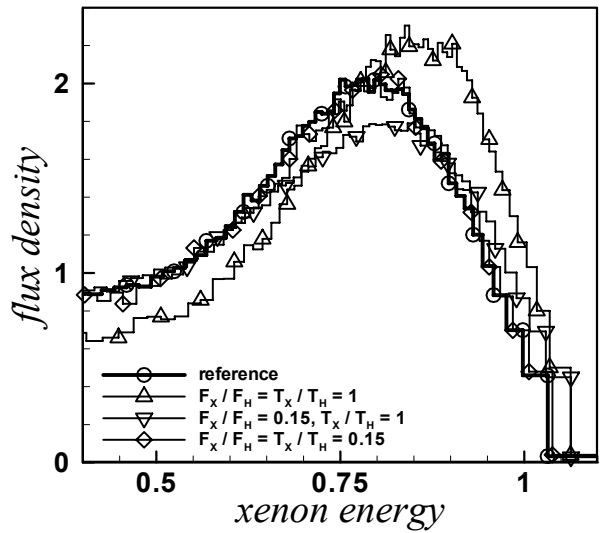 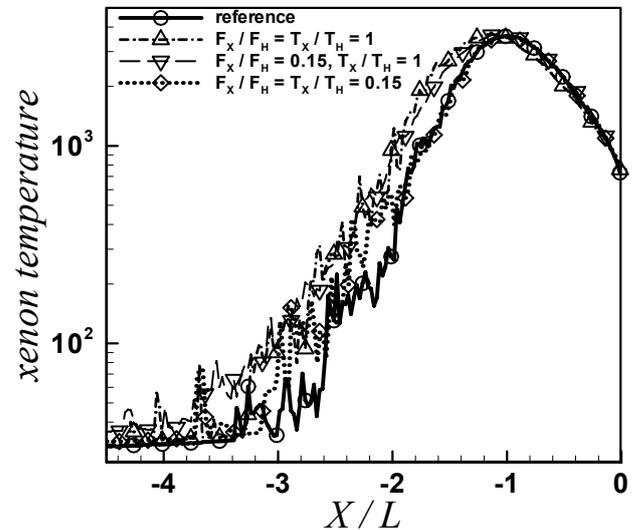

Fig. 3. High-energy part of energy spectrum of xenon simulators hitting the target, for simulations with different settings.

Fig. 4. Translational temperature profile of heavy component at the plane of symmetry, for simulations with different settings.

## 6 Simulating diffusion of silver nanoclusters in argon flow

After testing new approach on simple problems, it is time to demonstrate its use in conditions of real application. A good "baptism of fire" problem for the new approach would be the problem of silver nanocluster diffusion in transonic flow of argon carrier gas. This problem refers to numeric analysis of bactericide silver-fluoropolymer coatings deposition experiments [20] held in Kuteladze Institute of Thermophsyics SB RAS (Novosibirsk, Russia). Simulations are performed in conditions of one of routine experiments, where the source of fluoropolymer precursor is inactive, and only a flow from the silver vapor source contributes to deposition. As is known from experiments, in this case the coating is formed generally from silver nanoclusters formed inside the silver vapor source. Two fractions of silver particles are involved: small nanoclusters of few nanometers in diameter and nanoparticles of tens of nm.

Profiles of nanocluster flow in front of the surface for few fractions of nanoclusters are obtained from simulations. Experimental team have kindly agreed to set up additional experiment of silver deposition onto the fixed long narrow stainless steel plate, so the profile of the coating can be examined as well and compared with simulation results.

Axisymmetric problem setting is used in simulation. Standard radial spatial weighting scheme is used to level the number of simulators at different radii. Kannenberg's approach is used to get advantage on strong density difference, i.e. few regions of different cell size and time step are used. In the settling chamber, the cell is 30 μm and the time step is 40 ns. Outside the source, the cell size and time step are 3.5 times greater. The number of simulators is chosen so that $\frac{h}{\lambda \bar{N}} = 0.04$ near the silver surface, ~ 1 700 000 of argon simulators are present simultaneously in simulation domain.

Table 1. Model parameters for different components

| Component | VSS model parameters for collisions with argon, $T_{ref} = 950°C$ | | | Mass, a. u. | Internal degrees of freedom | Time step scaling factor |
| --- | --- | --- | --- | --- | --- | --- |
| | $D_{ref}$, Å | $\omega$ | $\alpha$ | | | |
| Ar | 3.28 | 0.65 | 1.28 | 40 | 0 | 1 |
| $Ag_{16}$ | 5.66 | 0.5 | 1 | 1728 | 87 | 0.33 |
| $Ag_{64}$ | 8.02 | 0.5 | 1 | 6912 | 375 | 0.15 |
| $Ag_{256}$ | 11.77 | 0.5 | 1 | 27648 | 1527 | 0.1 |
| $Ag_{1024}$ | 17.72 | 0.5 | 1 | 110592 | 6135 | 0.04 |

Nanocluster fractions of 16, 64, 256 and 1024 silver atoms are chosen for simulation. The concentration of nanoclusters is considered low enough not to disturb the argon and not to collide each other, i.e. component weights are used, the argon simulator velocity never change in argon-nanocluster collisions, thus, nanoclusters act like test particles. New approach is used as well: nanoclusters are slowed down up to 25 times. Hard sphere collision model is used for all Ar – nanocluster pairs, Borgnakke-Larsen model is used for internal degrees of freedom, with all collisions being non-elastic. The summary of molecular model parameters is shown in Table 1.

Fig. 5 shows the geometry of the problem setting, together with temperature field and streamlines of carrier gas (argon). The cylindrical settling chamber is on the left. Its left end is the surface of melt silver surface at 990ºC. The heated carrier gas with the temperature 940ºC is injected via 1 mm wide circular slit (6 evenly spaced holes in experiment) just above the silver surface. At $Z = -7$ mm, the crucible mug part is connected with the nozzle part. From $Z = -7$ mm and up to $Z = 0$ is the subsonic part of the nozzle, having the temperature of 890ºC. The small capillary at $Z = 0 – 1$ mm is 800ºC. From $Z = 1$ mm and up to $Z = 20.5$ mm is the diffuser of the nozzle, having temperature of 737ºC. The diameter of settling chamber is 13 mm, diameter of the nozzle critical section is 3 mm, and diameter of the nozzle exit is 23 mm. The wall at $Z = 20.5$ mm is the last heat screen, its temperature is set to 573ºC. Instead of the narrow plate target (about 1 cm in width), the disk of 23 mm in diameter is used in simulation, placed at $Z = 52$ mm and having temperature of 447ºC. $Z = 85$ mm is the right end of the computational domain, the diameter of computational domain is 100 mm.

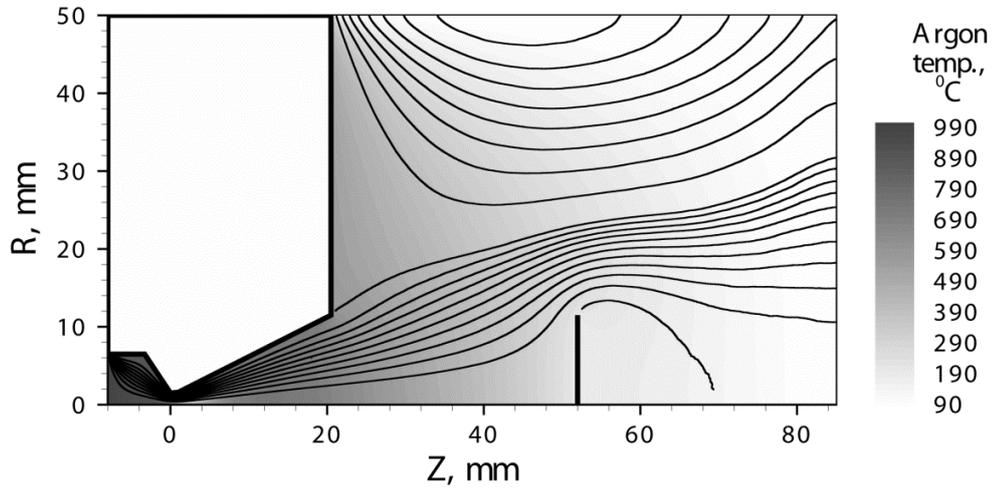

Fig. 5. Streamlines and temperature field of carrier argon flow

The pressure in the settling chamber is 2.1 torr, and the background pressure outside of the source is 0.14 torr. The background pressure was maintained in simulation by setting stream boundary conditions at the right and upper bounds, with given pressure, zero velocity, and temperature of 47ºC. One may see noticeable distortion of argon streamlines and temperature by these boundary conditions, yet, it does not ruin the purpose of simulation. The flow of argon is injected into the settling chamber, passes and leaves the supersonic nozzle, carries the background gas along, turns around the target, and leaves the computation domain. Streamlines are drawn so that argon flux between them is approximately the same (10%). At the plane of the target disk, the radius of a jet is about 23 – 25 mm. The vortex is formed behind the target.

Soon after reaching supersonic speed, the flow decelerates rather quickly and is subsonic in most part of the diffuser. No pronounced shock waves are formed, because of high rarefaction. Argon mean free path is ≈ 0.13 mm in the settling chamber, ≈ 0.2 mm in the critical section, and ≈ 0.6 mm at the target.

Now, the nanoclusters. In this simulation, nanoclusters are supposed to form heterogeneously on the surface of subsonic part of the nozzle, which have substantially lower temperature than the crucible with melt silver. They are "evaporated" from the surface, according to Maxwellian velocity distribution (like from inlet flow of zero velocity). While argon reflects diffusively from walls and target, nanoclusters are considered adsorbing on them, as well as on open bounds of computational domain. Rarefied argon flow is perceived much denser by larger nanoclusters. For example, the collision rate of $Ag_{1024}$ is 20 times greater in comparison with argon, and the mean free path of $Ag_{1024}$ in the settling chamber is less than 100 nm ($\approx$ 1000 times less than of argon). On the other hand, nearly 2000 collisions with argon are needed to lose 50% of momentum. The diffusion coefficient of $Ag_{1024}$ is about 50 times less in comparison to self-diffusion of argon (and about 35 times less than the diffusion coefficient for silver vapor).

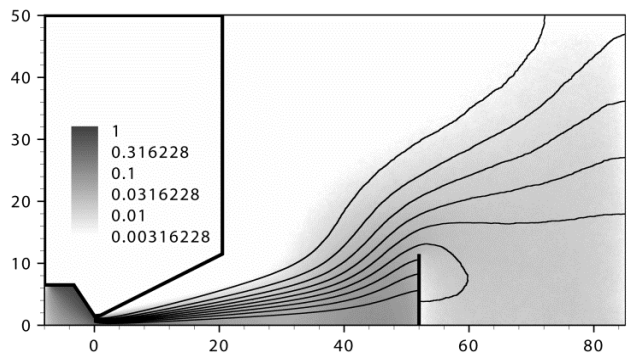 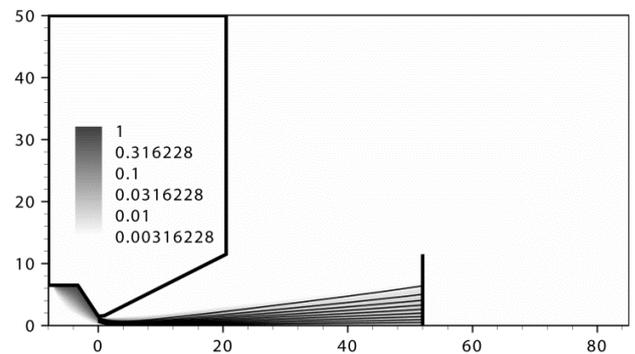

Fig. 6. Density field and streamlines of $Ag_{64}$ nanoclusters born in the settling chamber.

Fig. 7. Density field and streamlines of $Ag_{1024}$ nanoclusters born in the settling chamber.

Figure 6 shows the structure of the $Ag_{64}$ flow from the settling chamber. Streamlines are placed so that the flux between them is the same at $Z \approx 0$. Inside the chamber, streamlines are not shown, because mean velocity of nanoclusters is too small in comparison with statistical noise. This is the typical structure of nanocluster flow from the settling chamber. First, nanoclusters, emitted by subsonic part of the nozzle, diffuse into the argon boundary layer and are slowly carried towards the nozzle throat. Most of these nanoclusters will be adsorbed back onto emitting surface or somewhere else in the settling chamber. Remaining nanoclusters are accelerated up to very high speed (in comparison with their thermal velocity). Entering subsonic flow again, fast nanoclusters behave like macroscopic particles; they advance through the argon, gradually loosing speed with each collision. When they slow down to thermal velocity, the transition to diffusive mode happens ($Z \approx 35$ mm for $Ag_{64}$). Finally, nanoclusters suffer regular diffusion through argon like normal admixture molecules again. Some of them reach front side of the target; some diffuse behind it and reach its back side.

For $Ag_{16}$, the transition to diffusive mode happens at nozzle exit, for $Ag_{256}$, transition happens just in front of the target, and $Ag_{1024}$ (Fig. 7) hits the front side of the target before the transition happens. Only few simulators of $Ag_{1024}$ reached back side of the target.

Because of high mass, $Ag_{1024}$ do not decelerate down to thermal velocity (12 m/s) in front of the target, but only down to 105 m/s. Together with high internal heat capacity, this causes elevated internal temperature of nanoclusters when they hit the target. Latter might be important during bactericide metal-polymer film deposition process, because hot silver nanoclusters may chemically react with fluoropolymer on the target surface and thus affect bactericidal properties of the coating. However, one should also take into account very high surface energy of nanoclusters as well, especially of smaller ones; it may be released locally if a nanocluster merges with another one on the target surface.

Figure 8 shows flux density profiles of different nanoclusters in front of the target. As one can see, the heavier are nanoclusters, the higher is the flux at $R \approx 0$, but the narrower is the beam. Profiles for $Ag_{16}$ and $Ag_{64}$ are about as wide as the carrying jet of argon. However, beams of $Ag_{256}$ and, especially, $Ag_{1024}$ have rather sharp boundaries. While lighter two sorts of nanoclusters would deposit anywhere on the target front, heavier two would deposit only onto the central spot.

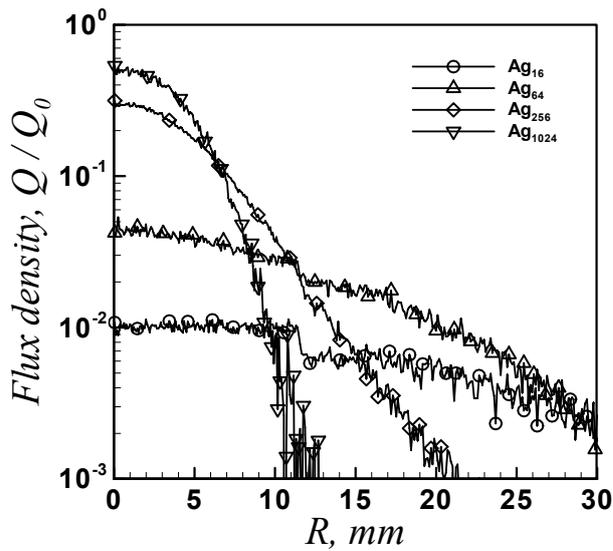
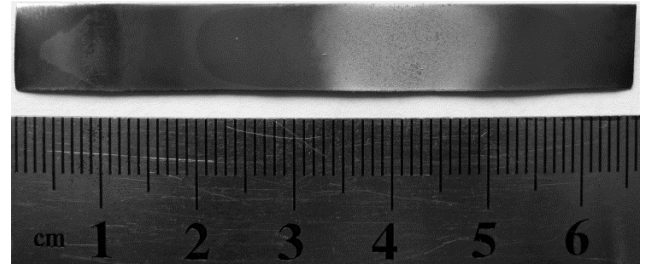

Fig. 8. Flux density profiles for different nanoclusters (born in the settling chamber only) at the plane of the target, referred to corresponding surface densities of emission.

Fig. 9. Photo of the target after deposition experiment.

Figure 9 presents a photo of target front side after deposition experiment. The diameter of white spot is in the good agreement with $Ag_{1024}$ profile from Fig. 8. This supports the adequacy of new method in modeling the diffusion of nanoclusters. Back side of the target stayed visually clear and unaffected (and was not photographed). Latter confirms that the majority of deposited silver comes from nanoclusters heavier than 256 atoms (otherwise, they would leave visible trace on the target back).

Note, only 1.7 million of argon simulators were used for all types of admixture nanoclusters, easy to fit in a memory of average workstation computer. To get the same precision without using new approach of slowing down nanoclusters, one would need over 40 million argon simulators to compute $Ag_{1024}$. Latter one would definitely be impractical for a workstation, as it would require gigabytes of memory.

Unfortunately, simulating diffusion of larger nanoparticles (tens of nanometers) with DSMC would be impractically expensive because of enormous collision frequency. Nevertheless, extrapolation of obtained results tells that nanoparticles would still form a narrow spot in the center of the target front, and not reach the back of the target.

## Conclusion

The new approach of using different time step for each component was proposed, with descriptions of algorithm changes and some recommendations on choosing appropriate component time steps and their effect on required number of simulators. This approach was tested on 1D heat transfer problem for He + Xe mixture, where helium was slowed down not to travel too much cells per time step, and on 1D and 2D variants of the problem of accelerated heavy Xe penetration through the compressed layer of light gas in front of an obstacle, where slowing down xenon allowed decreasing the number of helium simulators without drop in simulation precision. Finally, a complex 2D axisymmetric flow was simulated in conditions of real deposition experiment, where the coating was formed of silver nanoclusters born on inner surfaces of the silver vapor source and carried to the target by argon. The new approach made it possible without using enormous number of simulators. The obtained flux profile of $Ag_{1024}$ in front of the target turned out to be in good agreement with experimental results.

## Acknowledgements

Author is grateful to A. I. Safonov and other research team members from Institute of Thermophysics SB RAS for cooperation and providing experimental data used in section 6.

Author would be grateful for any support of his work, as he does not belong to any institution right now and have no sponsors. This would also help him keep his black hat in the closet.